\documentclass[11pt]{article}

\usepackage{geometry}
\usepackage{graphicx}
\usepackage{amsmath, amssymb, amsfonts, amsthm, float}  
\usepackage{enumerate, color, framed, float, multirow}
\usepackage{comment, longtable, caption, subcaption, appendix}
\usepackage[sort,longnamesfirst]{natbib}
\usepackage{setspace, parskip}
\usepackage{placeins}
\usepackage{makeidx}
\usepackage{fancyhdr}
\usepackage{bm}
\usepackage{mathtools}
\usepackage[bookmarks]{hyperref}
\hypersetup{
	bookmarks=true,         % show bookmarks bar?
	unicode=false,          % non-Latin characters in Acrobat's bookmarks
	pdftoolbar=true,        % show Acrobat's toolbar?
	pdfmenubar=true,        % show Acrobat's menu?
	pdffitwindow=true,      % page fit to window when opened
	pdftitle={My title},    % title
	pdfauthor={Author},     % author
	pdfsubject={Subject},   % subject of the document
	pdfnewwindow=true,      % links in new window
	pdfkeywords={keywords}, % list of keywords
	colorlinks=true,        % false: boxed links; true: colored links
	linkcolor=blue,         % color of internal links
	citecolor=blue,        % color of links to bibliography
	filecolor=magenta,      % color of file links
	urlcolor=cyan           % color of external links
}
\usepackage{booktabs}
\usepackage{framed}    % to make framed boxes
\usepackage{scalefnt}
\usepackage{alltt}
\usepackage{verbatim}
\usepackage{setspace}
\usepackage{footnote}
\usepackage{orcidlink}

\allowdisplaybreaks

\newcommand{\B}{\mathcal{B}}

\newcommand{\ind}{\perp\!\!\!\!\perp} 
\newcommand{\cH}{\mathcal{H}}
\newcommand{\cV}{\mathcal{V}}
\newcommand{\hatcV}{\hat{\mathcal{V}}}
\newcommand{\hatB}{\hat{\mathcal{B}}}

\newcommand{\df}{\mathrm{d}}

\newtheorem{theorem}{Theorem}[section]
\newtheorem{proposition}{Proposition}[section]
\newtheorem{lemma}{Lemma}[section]
\newtheorem{assumption}{Assumption}[section]

\usepackage{graphicx}

\title{A New Measure of Dependence Between Continuous and Multinomial Random Variables}
\author{Lu Yang\thanks{School of Statistics, University of Minnesota, email:luyang@umn.edu} \orcidlink{0000-0002-7538-6687} \and Galin L. Jones\thanks{School of Statistics, University of Minnesota, email:galin@umn.edu} \orcidlink{0000-0002-6869-6855}}

\begin{document}
\maketitle

\begin{abstract}
A novel measure of dependence between a continuous random variable and a multinomial random variable is introduced.  The proposed measure is based on the Hellinger distance between conditional distributions. It satisfies the desiderata for a dependence measure without making specific distributional assumptions about the continuous random variable or assuming that the discrete random variable arises from a latent continuous random variable.  An estimator of the dependence measure based on data splitting and kernel density estimation is developed. The asymptotic distribution of the estimator has a simple form with a convergence rate of $\sqrt{n}$, making confidence intervals for the dependence measure and a test for independence straightforward and computationally convenient.  
\end{abstract}

\noindent%
\vfill

\newpage

\section{Introduction}
\label{sec:intro}

The study of dependence between random variables is a fundamental task in statistics. In many applications, two random variables of interest differ in type, with one discrete and the other continuous.  For example, one might be interested in assessing the association between the use of electronic devices while studying and students' likelihood of passing an exam. Passing an exam is dichotomous, whereas the use of electronic devices can be measured on a continuum (e.g., total screen-viewing duration).  Other examples abound, but complications are encountered when applying established dependence measures to mixed-type random variables.  The focus here is on measuring the dependence between a continuous random variable and a multinomial random variable in a principled way. 

\begin{figure}
	\centering
        \includegraphics[width=0.4\textwidth]{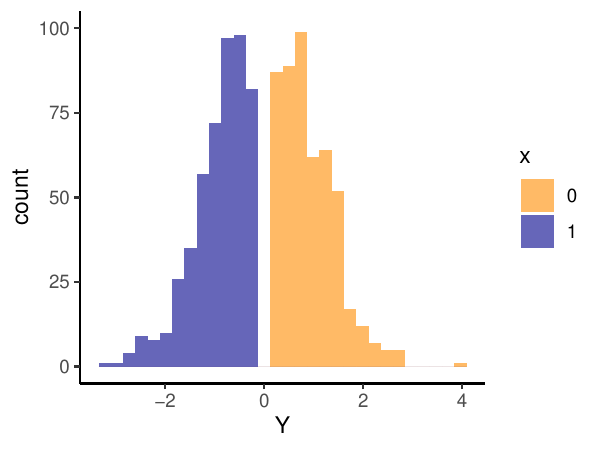}
        \includegraphics[width=0.4\textwidth]{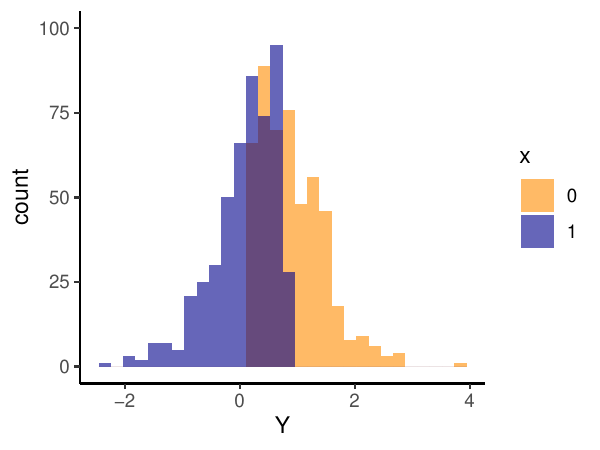}
	\caption{The left plot displays a histogram of 1000 simulated values under strict dependence of $X$ and $Y$. The right plot displays a histogram of 1000 simulated values when $X$ and $Y$ are associated, but not strictly dependent.  \label{fig:sep}}
\end{figure}

Suppose $X$ is a multinomial random variable and $Y$ is a continuous random variable. \cite{renyi1959measures} proposed seven criteria for a measure of dependence between any two random variables.  However, some of R\'enyi's original criteria are inappropriate for the current setting or simply too strong. For example, the requirement that the dependence measure be defined for all random variables is not satisfied even for standard measures of dependence such as Pearson's correlation.  Other criteria, such as symmetry, are inappropriate here. Adapting R\'enyi's criteria, a measure of dependence between $X$ and $Y$, denoted $\cV(X,Y)$, should satisfy the following desiderata: 
\begin{enumerate}
\item[(A)] $0\leq \cV(X,Y) \leq 1$;
\item[(B)] $\cV(X,Y)=0$ if and only if $X$ and $Y$ are independent;
\item[(C)] $\cV(X,Y) = 1$ if and only if there is a strict dependence between $X$ and $Y$; 
\item[(D)] if the Borel measurable function  $g: \mathbb{R} \to \mathbb{R}$ is one-to-one and onto, $\cV(X,g(Y))=\cV(X,Y)$; and
\item[(E)] $\cV(X, Y)$ does not depend on the encoding of the category labels for $X$.
\end{enumerate} 
Strict dependence exists between the random variables $X$ and $Y$ when the value of $Y$ determines the value of $X$. For simplicity, consider the case where $X$ is binary.  Then strict dependence is equivalent to an empty intersection of the supports of $Y \mid X=0$ and $Y \mid X=1$; this extends naturally to the more general multinomial setting.  Figure~\ref{fig:sep} illustrates this notion of strict dependence in the binary setting.

Existing dependence measures for mixed-type data fail to satisfy the desiderata above  and often require strong distributional assumptions. Dependence measures have been proposed in the case where $X$ is binary and $Y$ is Gaussian  \citep{tate1955theory}. It assumes the existence of a latent variable $Z$ such that $X=1(Z>w)$ with a threshold $w$  and $(Y, Z)$ follows a bivariate normal distribution with correlation $\rho$, known as the biserial correlation  
\citep{cox1974estimation, kraemer1981modified, pearson1909new}. \cite{olki:tate:1961} extended Tate's approach to the more challenging multinomial setting. 
\cite{chen:wehr:2014} proposed a method based on generalized linear mixed models. %and also imposed strong distributional assumptions.  
However, these approaches rely on distributional assumptions, limiting their applicability and potentially leading to misleading conclusions when these assumptions are not met.

Dependence measures often are designed for continuous outcomes, and their application to mixed-type data can be problematic. Classical measures such as Pearson correlation coefficients and Kendall's tau can be misleading when applied to mixed-type data, yielding values near zero even when the random variables are associated \citep{josse2016measuring}, thereby violating desideratum (B) and potentially ruling out important associations. Measures based on ranks or nearest neighbors \citep{hoeffding1994non,chatterjee2021new, han2021extensions} require an ordering of the support and thus assigning numerical values to a multinomial outcome. As a result, the strength of dependence depends on the arbitrary encoding of a multinomial outcome, which violates desideratum (E). Furthermore, these methods break ties in discrete data randomly, which makes interpretation challenging. Rank-based approaches are closely related to copula-based methods  \citep[see, e.g.,][]{dette2013copula, geenens2022hellinger}, which are based on cumulative distribution functions and thus also require the ordering in the support.  \cite{szekely2007measuring} proposed distance correlation as the weighted distance between the joint and the product of the marginal characteristic functions of random variables.  More recently, \cite{zhong2023semi} extended distance correlations to handle mixed-type data. However, while distance correlation and its variations can achieve a value of zero under independence, it cannot attain the upper bound of 1 in the presence of strict dependence \citep{szekely2023energy}, failing to satisfy desideratum (C).

There is a large body of literature on independence testing for general types of outcomes \citep[see, e.g.,][]{berrett2019nonparametric,shi2022distribution}, many based on the discrepancy between the joint distribution and the product of marginal distributions \citep[see, e.g.,][]{hoeffding1994non, gretton2005kernel, szekely2007measuring} or partitioning the sample space \citep{heller2016consistent, zhang2019bet}.
However, these works do not focus on measuring the strength of dependence.

We propose a new measure of dependence between a continuous random variable and a multinomial random variable, based on the Hellinger distance between conditional distributions.  The measure satisfies the desiderata, which, to our knowledge, makes it the first such measure for mixed-type data to do so. Distributional assumptions are not required for the continuous random variable; there is no assumption that the discrete random variable arises from an underlying continuous random variable; and the measure does not depend on the encoding of the multinomial variable. 
We develop an estimator of the dependence measure
based on data splitting and kernel density estimation. 
The asymptotic distribution of the estimator has a simple form, facilitating computationally efficient inference including a confidence interval and a test for independence
without the need for the bootstrap. %and enabling the development of an interval estimator and a test for independence.  
Moreover, the estimator converges at a $\sqrt{n}$ rate despite kernel density estimation is
employed.

The remainder is organized as follows.  Section~\ref{sec:proposed} focuses on the setting where the discrete random variable is Bernoulli, and introduces the proposed dependence measure, its estimation, and asymptotic distribution.
This is generalized to the multinomial setting in Section~\ref{sec:ext}.  Illustrative examples and  simulation experiments are considered in both Section~\ref{sec:proposed} and~\ref{sec:ext}.  Two data applications are given in Section~\ref{sec:applications}.  Section~\ref{sec:final} contains some final remarks.  All proofs are deferred to the supplementary material.

\section{Binary and Continuous Variables}
\label{sec:proposed}
Suppose $X$ is Bernoulli and $Y$ is continuous; the generalization to the multinomial setting is handled in Section~\ref{sec:ext}.  If $X$ and $Y$ are independent, the conditional densities $f_{Y\mid X=0}$ and $f_{Y \mid X=1}$ are identical, while strict dependence corresponds to scenarios where $f_{Y \mid X=0}$ and $f_{Y \mid X=1}$ have different density supports.  Therefore, the discrepancy between $f_{Y \mid X=0}$ and $f_{Y \mid X=1}$ reveals the strength of dependence between $X$ and $Y$.  

Consider the Hellinger distance between $f_{Y \mid X=0}$ and $f_{Y \mid X=1}$, denoted $\mathcal{H} \left(f_{Y \mid X=0}, f_{Y \mid X=1}\right)$, and define a measure of dependence
\begin{align}
\label{eq:binary VH}
\begin{split}	
\mathcal{V}_H(X,Y) := & \mathcal{H}\left(f_{Y \mid X=0},f_{Y \mid X=1}\right) \\
= & \left[{\frac{1}{2}\int\left(\sqrt{f_{Y \mid X=0}(y)}-\sqrt{f_{Y \mid X=1}(y)}\right)^2\df y}\right]^{1/2}\\
= &
		\left[{1-\int\sqrt{f_{Y \mid X=0}(y)f_{Y \mid X=1}(y)}\df y}\right]^{1/2}.
\end{split}
\end{align} 
The term
\begin{align*}
%\label{eq:b}
%\mathcal{B}\left(f_{Y \mid X=0}, f_{Y \mid X=1}\right) 
\mathcal{B}_{01}:= \int \sqrt{f_{Y \mid X=0}(y) f_{Y \mid X=1}(y)} \df y
\end{align*} 
is the so-called Bhattacharyya coefficient \citep{bhat:1946} which is closely related to the R{\' e}nyi  divergence \citep{krishnamurthy2014nonparametric}.
While the Hellinger distance has been used to measure and test for dependence \citep{su2008nonparametric,geenens2022hellinger}, it is employed in a novel way here.   For instance, the dependence measure in \cite{geenens2022hellinger} is based on copulas and designed specifically for continuous outcomes.

Compared to many other measures of distributional similarity (e.g.,
energy distance \citep{szekely2023energy}, maximum mean discrepancy \citep{gretton2012kernel}, and Kolmogorov–Smirnov and Cramér–von Mises statistics), the use of Hellinger distance ensures that $\cV_{H}(X,Y)$ satisfies the desiderata; see Appendix for the details.  Moreover, the computational requirements for Hellinger distance are less than those of some other measures, such as total variation.

\subsection{Estimation}
\label{sec:esti}

Estimation of  $\mathcal{V}_H(X,Y)$ involves  two stages: (i) estimate the densities  $f_{Y \mid X=0}$  and $f_{Y \mid X=1}$ to obtain $\hat{f}_{Y \mid X=0}$ and $\hat{f}_{Y \mid X=1}$, respectively, and (ii) calculate the integral 
\begin{equation}
\label{eq:estimated b}
    \int\sqrt{\hat{f}_{Y \mid X=0}(y)\hat{f}_{Y \mid X=1}(y)}~\df y
\end{equation} to estimate the Bhattacharyya coefficient, and subsequently obtain the estimate for $\mathcal{V}_H(X,Y)$. The two steps are addressed with data splitting. Consider independent and identically distributed random variables $(X_i,Y_i)_{i=1}^n$ randomly split into $(X_i,Y_i)_{i=1}^{ \lfloor n/2 \rfloor}$  and $(X_i,Y_i)_{i=\lfloor n/2 \rfloor + 1}^n$, with a slight abuse of notation. The first half is used to estimate the densities in step (i) and the second half to estimate the integral in step (ii).  The details are now described.

In step (i), consider using a kernel density estimator (KDE) to obtain  $\hat{f}_{Y \mid X=0}$ and $\hat{f}_{Y \mid X=1}$.  Specifically, if $K$ denotes the kernel and $h_j$, $j=0,1,$ are bandwidths, possibly chosen by cross-validation  \citep{li2023nonparametric}, the estimators are given by
\begin{equation*}
\begin{split}
    {\hat {f}}_{Y \mid X=j}(y) & =
	{\frac {1}{h_j\sum_{i=1}^{\lfloor n/2 \rfloor }1\left(X_i=j\right)}}\sum _{i=1}^{\lfloor n/2 \rfloor }1(X_i=j)K{\Big (}{\frac {y-Y_{i}}{h_j}}{\Big )} .
\end{split}
\end{equation*}
The use of a KDE accommodates scenarios with bounded or unbounded support, and areas outside the support are automatically assigned a value of zero \citep{cuevas1997plug}. This is advantageous since the density supports are informative for the strength of dependence.

Consider step (ii), that is, estimating the integral in \eqref{eq:estimated b}. Following \cite{krishnamurthy2014nonparametric}, a first-order Von Mises expansion (VME) of $\mathcal{B}_{01}$ at $({f}_{Y \mid X=0},  {f}_{Y \mid X=1})$ is used to correct for the discrepancies between the estimated densities and their estimands, yielding
\begin{align}\label{eq:plug}
	\begin{split}
	\mathcal{B}_{01}=&\frac{1}{2}\int\sqrt{\frac{\hat{f}_{Y \mid X=0}(y)}{\hat{f}_{Y \mid X=1}(y)}}{f}_{Y \mid X=1}(y)\df y+\frac{1}{2}\int\sqrt{\frac{\hat{f}_{Y \mid X=1}(y)}{{\hat{f}_{Y \mid X=0}(y)}}}{f}_{Y \mid X=0}(y)\df y+R\\=&
	\frac{1}{2}\mathrm{E}_{Y \mid X=1}
	\sqrt{\frac{\hat{f}_{Y \mid X=0}(Y)}{\hat{f}_{Y \mid X=1}(Y)}}+\frac{1}{2}\mathrm{E}_{Y \mid X=0}\sqrt{\frac{\hat{f}_{Y \mid X=1}(y)}{{\hat{f}_{Y \mid X=0}(y)}}}+R,\end{split}
\end{align}
where $R$ is the remainder.  Replacing the expectation in \eqref{eq:plug} with the sample average,  the proposed estimator of the Bhattacharyya coefficient is 
\begin{align}
\label{eq:split}
\begin{split}
%\hat{\mathcal{B}}(\hat{f}_{Y \mid X=0},\hat{f}_{Y \mid X=1}) 
\hat{\mathcal{B}}_{01}= & \frac{1}{2\sum_{i=\lfloor n/2 \rfloor +1}^n1 \left(X_i=0\right)} \sum_{i= \lfloor n/2 \rfloor +1}^n1 \left(X_i=0\right)\sqrt{\frac{\hat{f}_{Y \mid X=1}(Y_i)}{\hat{f}_{Y \mid X=0}(Y_i)}}\\
    &~+\frac{1}{2\sum_{i= \lfloor n/2 \rfloor +1}^n1 \left(X_i=1\right)} \sum_{i=\lfloor n/2 \rfloor +1}^n1\left(X_i=1\right)\sqrt{\frac{\hat{f}_{Y \mid X=0}(Y_i)}{\hat{f}_{Y \mid X=1}(Y_i)}}.
\end{split}
\end{align}
Finally, the proposed estimator for the dependence measure is  $\hat{\mathcal{V}}_H(X,Y)=\sqrt{{1-\hat{\mathcal{B}}_{01}}}.$

It is natural to generalize the data splitting method described above to $M$-fold estimation.  In this case the data is split into $M$ equal-sized subsets $\mathcal{I}_1,\ldots\mathcal{I}_M$. 
For each $m\in \{1,\ldots,M\}$, the  folds of data $\mathcal{I}_1,\ldots,\mathcal{I}_{m-1},\mathcal{I}_{m+1},\ldots,\mathcal{I}_M$ are used to estimate the densities $\hat{f}_{Y|X=0}$ and $\hat{f}_{Y|X=1}$. Then, the required integral \eqref{eq:estimated b} is estimated based on $\mathcal{I}_{m}$:
 \begin{align*}
\hat{\mathcal{B}}_m & :=	\frac{1}{2\sum_{i\in\mathcal{I}_m}1\left(X_i=0\right)}\sum_{i\in\mathcal{I}_m}1\left(X_i=0\right)\sqrt{\frac{\hat{f}_{Y|X=1}(Y_i)}{\hat{f}_{Y|X=0}(Y_i)}} \\
& \quad + \frac{1}{2\sum_{i\in\mathcal{I}_m}1\left(X_i=1\right)}\sum_{i\in\mathcal{I}_m}1\left(X_i=1\right)\sqrt{\frac{\hat{f}_{Y|X=0}(Y_i)}{\hat{f}_{Y|X=1}(Y_i)}}.\end{align*}
The $M$-fold  estimator of $\mathcal{B}_{01}$ is 
\begin{align}
\label{eq:ave}
\hat{\mathcal{B}}_{01}^{M-fold} =\frac{1}{M}\sum_{m=1}^{M}\hat{\mathcal{B}}_m.
\end{align}

\subsubsection{Illustrative examples}
\label{sec:simudemo}

%The estimator is demonstrated, and its performance is compared with several other measures in the following scenarios.
To immediately  demonstrate   the advantages of the proposed dependence measure $\cV_{H}(X,Y)$, we compare its estimate with several other estimated measures in the following scenarios.
\begin{enumerate}
	\item  Suppose $X \sim \mathrm{Bernoulli}(0.5)$,
	$Y \mid X=0\sim \mathrm{Uniform}(-2,-1)$, and $Y \mid X=1\sim \mathrm{Uniform}(0,1)$. Notice that the supports of $Y \mid X=0$ and $Y \mid X=1$ do not overlap and are well separated.  As such, this is a clear case of strict dependence.
	\item  Suppose $X \sim \mathrm{Bernoulli}(0.5)$,  $Y\mid X=0\sim \mathrm{Uniform}(-1,0)$, and $Y \mid X=1\sim \mathrm{Uniform}(0,1)$.  As in scenario 1, the supports of $Y \mid X=0$ and $Y \mid X=1$ do not overlap, but are adjacent.  This is again a case of strict dependence, but perhaps less clearly than in the first scenario. 
	\item Suppose $X \sim \mathrm{Bernoulli}(0.5)$, $Y \mid X=0\sim \mathrm{N}(0,1)$, and $Y \mid X=1\sim \mathrm{N}(-1,1)$. Equivalently, this is a linear regression model  $Y \mid X\sim N(-X,1)$, so this illustrates the case of a difference in mean and linear dependence.
	\item  Suppose $X \sim \mathrm{Bernoulli}(0.5)$, $Y \mid X=0\sim N(1,1)$, and $Y \mid X=1\sim N(1,2^2)$, so this is a case of a difference in variance.
	\item Suppose $X \mid Y \sim \mathrm{Bernoulli}(P(X=1 \mid Y))$ with 
    \[
    P(X=1 \mid Y)=\text{logit}^{-1}\left(-2 Y1(Y<0)+2 Y1(Y>0)\right)
    \]
    and $Y \sim \mathrm{N}(0,1)$. Thus, this is a case of nonlinear dependence. 
	\item  Suppose $X \sim \mathrm{Bernoulli}(0.5)$ and $Y \sim \mathrm{N}(0,1)$ are independent.  
\end{enumerate}

\begin{figure} \centering
	\includegraphics[width=.9\textwidth]{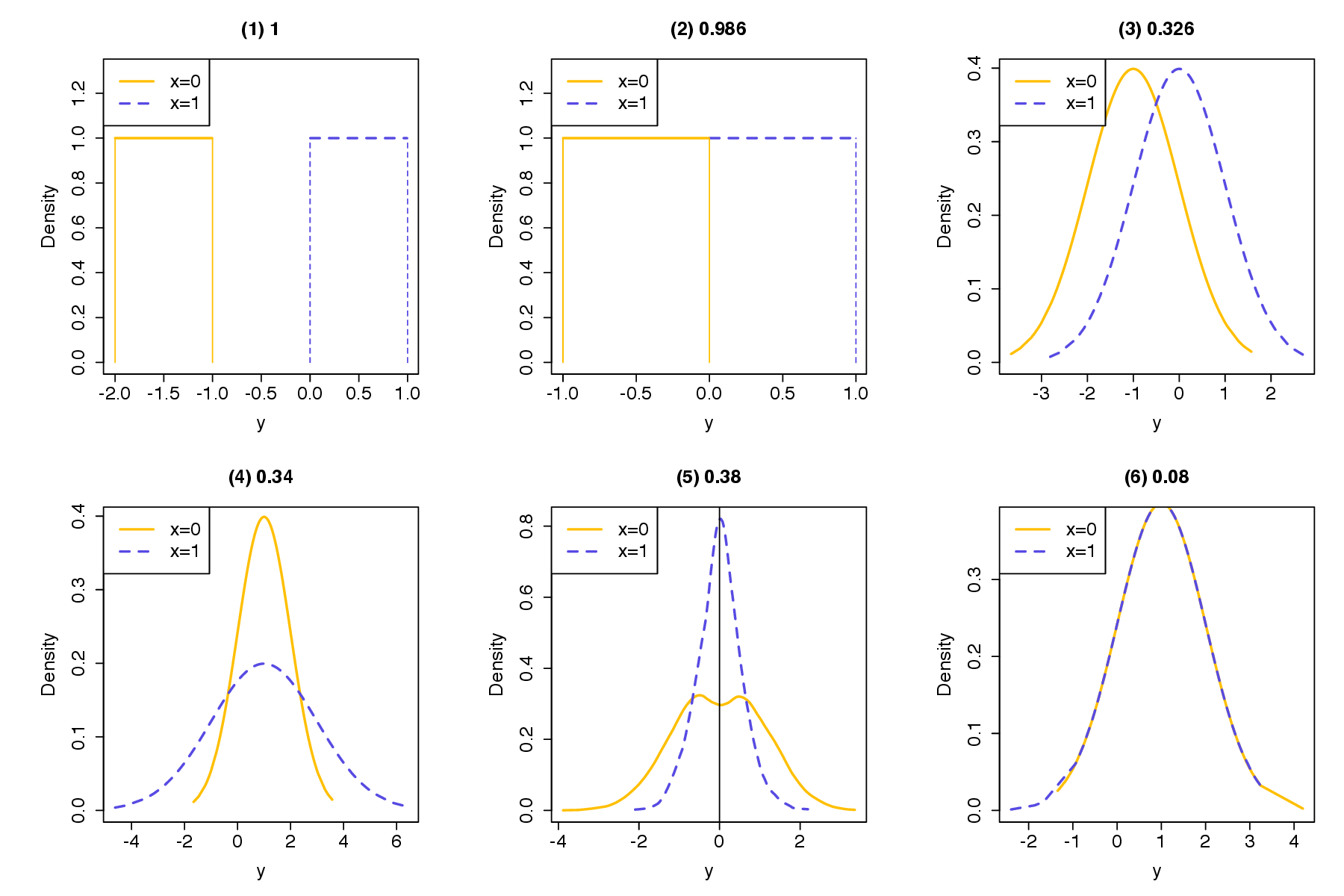}\vspace{-1em}
	\caption{Values of $\hat{\mathcal{V}}_H(X, Y)$ %along with $\hat{f}_{Y \mid X=0}$ and $\hat{f}_{Y \mid X=1}$ 
    in six scenarios. Sample size 500.\label{fig:demo}}
\end{figure}

The scenarios are displayed in Figure~\ref{fig:demo}.  
In each of the six scenarios, independent pairs $(X_i, Y_i)$ for $i=1,\ldots, 500,$ are generated, and the value of $\hatcV_{H}(X,Y)$ is calculated. %along with the estimated densities $\hat{f}_{Y \mid X=0}$ and $\hat{f}_{Y \mid X=1}$.  

Table~\ref{tab:other} includes estimates of $\cV_{H}(X, Y)$ along with estimates of the  Pearson, Kendall,  Spearman, and distance correlation coefficients as well as the   semi-distance correlation \citep{zhong2023semi} which was designed for mixed data. In scenario 1, a case of strict dependence, $\hatcV_{H}(X, Y) = 1$, demonstrating that it satisfies desideratum (C).  Note that none of the alternative dependence measures achieves this.  In scenario 2, again a case of strict dependence but with a weaker signal, $\hatcV_{H}(X, Y)$ is close to 1 and is larger than the alternative measures which are noticeably smaller than 1. In the linear dependence of scenario 3, all dependence measures are comparable, except the semi-distance correlation, which is much smaller. However, when the variances differ, or when there is nonlinear dependence, as in scenarios 4 and 5, respectively, other  measures fail to detect dependence, especially the Pearson, Kendall, Spearman, and semi-distance correlation coefficients; see the discussions in \cite{szekely2007measuring} and \cite{josse2016measuring}.  When the variables are independent, as in scenario 6, all of the dependence measures are close to 0. 

\begin{table} \centering 
	\caption{Estimated dependence measures in six scenarios.
		\label{tab:other} }
	\begin{tabular}{@{\extracolsep{5pt}} ccccccc} 
		\toprule
		Scenario& $\hat{\mathcal{V}}_H(X, Y)$ & Pearson & Kendall & Spearman & Distance &Semi-distance\\ 
		\midrule
% sample size 1000        
%		1 & 1 & 0.962 & 0.707 & 0.866 & 0.978 \\ 
%		2 & 0.994 & 0.869 & 0.707 & 0.866 & 0.893 \\ 
%		3 & 0.370 & 0.462 & 0.381 & 0.466 & 0.453 \\ 
%		4 & 0.333 & 0.024 & 0.028 & 0.034 & 0.221 \\ 
%		5 & 0.324 & 0.023 & 0.018 & 0.022 & 0.161 \\ 
%		6 & 0.071 & -0.017 & -0.013 & -0.016 & 0.045 \\ 
%1 & $1$ & $0.960$ & $0.707$ & $0.864$ & $0.977$ \\ 
%2 & $0.986$ & $0.861$ & $0.707$ & $0.864$ & $0.884$ \\ 
%3 & $0.326$ & $0.416$ & $0.340$ & $0.416$ & $0.403$ \\ 
%4 & $0.340$ & $$-$0.032$ & $$-$0.013$ & $$-$0.015$ & $0.234$ \\ 
%5 & $0.380$ & $$-$0.035$ & $$-$0.040$ & $$-$0.048$ & $0.191$ \\ 
%6 & $0.080$ & $0.016$ & $0.025$ & $0.031$ & $0.048$ \\ 
1 & $1$ & $0.960$ & $0.707$ & $0.864$ & $0.977$ & $0.954$ \\ 
2 & $0.986$ & $0.861$ & $0.707$ & $0.864$ & $0.884$ & $0.782$ \\ 
3 & $0.326$ & $0.416$ & $0.340$ & $0.416$ & $0.403$ & $0.163$ \\ 
4 & $0.340$ & $$-$0.032$ & $$-$0.013$ & $$-$0.015$ & $0.234$ & $0.055$ \\ 
5 & $0.380$ & $$-$0.035$ & $$-$0.040$ & $$-$0.048$ & $0.191$ & $0.036$ \\ 
6 & $0.080$ & $0.016$ & $0.025$ & $0.031$ & $0.048$ & $0.002$ \\ 
		\bottomrule
	\end{tabular} 
\end{table}

\subsection{Large Sample Properties}
\label{sec:theo}
The asymptotic distribution of $\mathcal{V}_H(X, Y)$ is established under non-independence in Section~\ref{sec:largesample_dep} and independence in Section~\ref{sec:test}, respectively. The following standing assumption is required throughout.

\begin{assumption}
\label{assum:1}
The second-order derivatives of $f_{Y \mid X=j}$,  for $j=0,1$, and their integrals are bounded.
\end{assumption}

\subsubsection{Asymptotic Distribution under Non-Independence}
\label{sec:largesample_dep}

The asymptotic distribution is developed in Theorem~\ref{theo:nonind} and confidence intervals for the dependence measure are discussed.  Let $p_0=\Pr(X=0)$ and $p_1=\Pr(X=1)$.

\begin{theorem}
\label{theo:nonind}
Let $v= \frac{1}{2p_0}+\frac{1}{2p_1}$. Under Assumption~\ref{assum:1}, if $X$ and $Y$ are not independent, then, as $n \to \infty$,
	$$\sqrt{n} \left(\hatB_{01} - \B_{01} \right) \rightarrow_d {\rm N} \left(0,{  v \left(1-\B_{01}^2\right)} \right).$$
\end{theorem}
Proofs of the theoretical results are deferred to the the supplementary material.
As an immediate result of Theorem \ref{theo:nonind}, a confidence interval for $\B_{01}$ and $\mathcal{V}_H(X, Y)$ with a simple form can be constructed, which is outlined below. Thus, the bootstrap is not required for inference at any point. Note also that despite employing KDE, a $\sqrt{n}$ convergence rate is achieved. This is because step (ii) of the estimation is based on a sample average. The same rate has been obtained for other data-splitting estimators involving the KDE \citep[e.g.,][]{bickel1988estimating}. Our results should not be confused with those obtained when the Hellinger distance was used to measure the discrepancy between the true and estimated densities, which have a slower convergence rate \citep{kanazawa1993hellinger}.
 
 Let $\hat{p}_0$ and $\hat{p}_1$ be the usual empirical estimates of $p_{0}$ and $p_{1}$, respectively, and let
$\hat{v} = \frac{1}{2\hat{p}_0}+\frac{1}{2\hat{p}_1}$.
Then, if $z_{1-\alpha/2}$ denotes the $(1- \alpha/2)$th standard normal quantile, an approximate $100(1-\alpha)\%$ confidence interval for $\B_{01}$ is given by
\begin{align}\label{eq:cib}
	\hatB_{01} \pm z_{1-\alpha/2} \left[ \frac{ \hat{v} \left(1-\hatB_{01}^{2}\right)}{n} \right]^{1/2}.
\end{align}

There are two straightforward approaches to
 constructing a $100(1-\alpha)\%$ confidence interval for $\mathcal{V}_H(X,Y)$ subsequently.
 Based on  a monotone transformation, an approximate confidence interval for $\mathcal{V}_H(X,Y)$ is 
 \begin{align}
     	\label{eq:ci}
 \left\lbrace1-\hatB_{01} \pm z_{1-\alpha/2} \left[ \frac{ \hat{v} \left(1-\hatB_{01}^{2}\right)}{n} \right]^{1/2}\right\rbrace^{1/2}.
 \end{align}
An alternative is  obtained via the delta method, 
\begin{equation*}
   % \label{eq:delta method CI}
    \hat{\mathcal{V}}_{H} \pm z_{1-\alpha/2} \left[ \frac{\hat{v} (1 + \hatB_{01})}{4n} \right]^{1/2}.
\end{equation*}
In our empirical work, these confidence intervals performed similarly, with a slight edge in coverage for the interval in expression \eqref{eq:ci} and hence it is the one used in the remainder.

Theorem~\ref{theo:nonind} can be extended to the $M$-fold estimation setting.
\begin{proposition}
\label{prop:cv}
Under Assumption \ref{assum:1}, when $X$ and $Y$ are not independent, as $n \to \infty$, 
\begin{equation*}
    \sqrt{n} \left(\hatB_{01}^{M-fold} - \B_{01} \right) \rightarrow_d  {\rm N} \left(0, \frac{v}{2} \left(1-\B_{01}^2\right) \right).
\end{equation*}

\end{proposition}

Comparing Proposition~\ref{prop:cv} to Theorem~\ref{theo:nonind} shows that using $M$-fold estimation reduces the asymptotic variance by half.  Of course, this comes at the expense of an often modest increase in computational effort. 

\subsubsection{Asymptotic Distribution under Independence and Independence Test}
\label{sec:test}

Under independence, the estimator converges at a faster rate. This is because the key term contributing to the asymptotic distribution under non-independence degenerates under independence, necessitating the VME of higher orders. For simplicity, we use the same bandwidth $h$ for estimating $\hat{f}_{Y \mid X=0}$ and $\hat{f}_{Y \mid X=1}$ in this section,   as the two distributions are similar under independence, though the results can be extended to allow different bandwidths. Next, recall that $K$ denotes the kernel for the KDE and define
\begin{equation*}
 \sigma_K^2 = \int K^2(y) \df y \qquad \mathrm{and} \qquad \kappa_K^2 = \int\left(\int K\left(u\right)K\left(u+z\right) \df u  \right)^2\df z  . 
\end{equation*}
Both integrals exist for commonly used kernels and are often straightforward to calculate. 

\begin{theorem}
\label{theo:ind}
	If $X$ and $Y$ are  independent and $S_0 < \infty$ is the size of the support, then, as $n \to \infty$,
	\begin{align*}
		\sqrt{n^2h}\left(\hatB_{01}-1-\frac{S_0\sigma_K^2}{4nhp_0p_1}\right)\rightarrow_d N\left(0,
		\frac{S_0}{p_0^2p_1^2}\left(\frac{\kappa_K^2}{8}+\frac{\sigma_K^2}{4}\right)\right).
	\end{align*}
\end{theorem}

The convergence rate under independence is $\sqrt{n^2h}$. Under the optimal bandwidth of KDE which is $O(n^{-1/5}),$ this is $n^{9/10}$, faster than $\sqrt{n}$.
%It  implies that under independence, $\hatB_{01}-1-\frac{S_0\sigma_K^2}{4nhp_0p_1}$ converges to 0 faster than $\sqrt{n}$, or equivalently, $\sqrt{n}\left(\hatB_{01}-1-\frac{S_0\sigma_K^2}{4nhp_0p_1}\right)\rightarrow_p0$. 

Making use of Theorem~\ref{theo:ind} requires the value, or at least an estimate, of $S_0$. One  approach is transforming $Y$ using its empirical distribution function, before estimating $\cV_{H}(X,Y)$. This yields $S_{0} = 1$ and, by property (D) of the desiderata, does not change the value of $\cV_{H}(X,Y)$.

Testing for independence between $X$ and $Y$ is equivalent to testing 
\begin{equation*}
    H_0:\mathcal{V}_H(X,Y)=0 \qquad \mathrm{versus} \qquad H_A:\mathcal{V}_H(X,Y)>0.
\end{equation*} 

Using Theorem~\ref{theo:ind} yields a test statistic 
\begin{equation*}
    T_n = \frac{\sqrt{n^2h}\left(\hatB_{01} - 1-  \frac{\sigma_K^2S_0}{4nh\hat{p}_0\hat{p}_1}\right)}{\sqrt{\frac{S_0\left(\kappa_K^2 + 2\sigma_K^2\right)}{8 \hat{p}_0^2\hat{p}_1^2}}} .
\end{equation*}
If $\Phi$ denotes the standard normal distribution function, the closed-form p-value is given by $$2\left(1-\Phi\left(|T_n|\right)\right).$$

   Under the null hypothesis, the test statistic is stochastically bounded and converges to a standard normal distribution. Under the alternative hypothesis, 
	by Theorem~\ref{theo:nonind}, $\hatB_{01}$ converges to a normal distribution with a rate of $\sqrt{n}.$ Therefore, under the alternative, the test statistic converges to $\infty$ and the p-value converges to 0, guaranteeing the power of the test.
    
\subsection{Simulation Experiments}
\label{sec:simu}

\subsubsection{Estimation}
\label{sec:simuest}

The finite-sample performance of the estimators \eqref{eq:split} and \eqref{eq:ave} is demonstrated in a setting where the degree of dependence can be adjusted.  If $X \sim \mathrm{Bernoulli}(0.4)$ and $Y \mid X=0 \sim \mathrm{N}(1,1)$ and $Y \mid X=1 \sim \mathrm{N}(1,\sigma^2)$, then 
$\B_{01} = \sqrt{{2\sigma}/{(1 + \sigma^2)}}.$ 
If $\sigma \in \{1.5, 2, 3\}$, then $\B_{01}$ is 0.961, 0.894, and 0.775, and $\cV_H(X, Y)$ is 0.198, 0.325, and 0.616, under weak, moderate, and strong dependence, respectively.

\begin{figure}
	\centering
	\includegraphics[width=\textwidth]{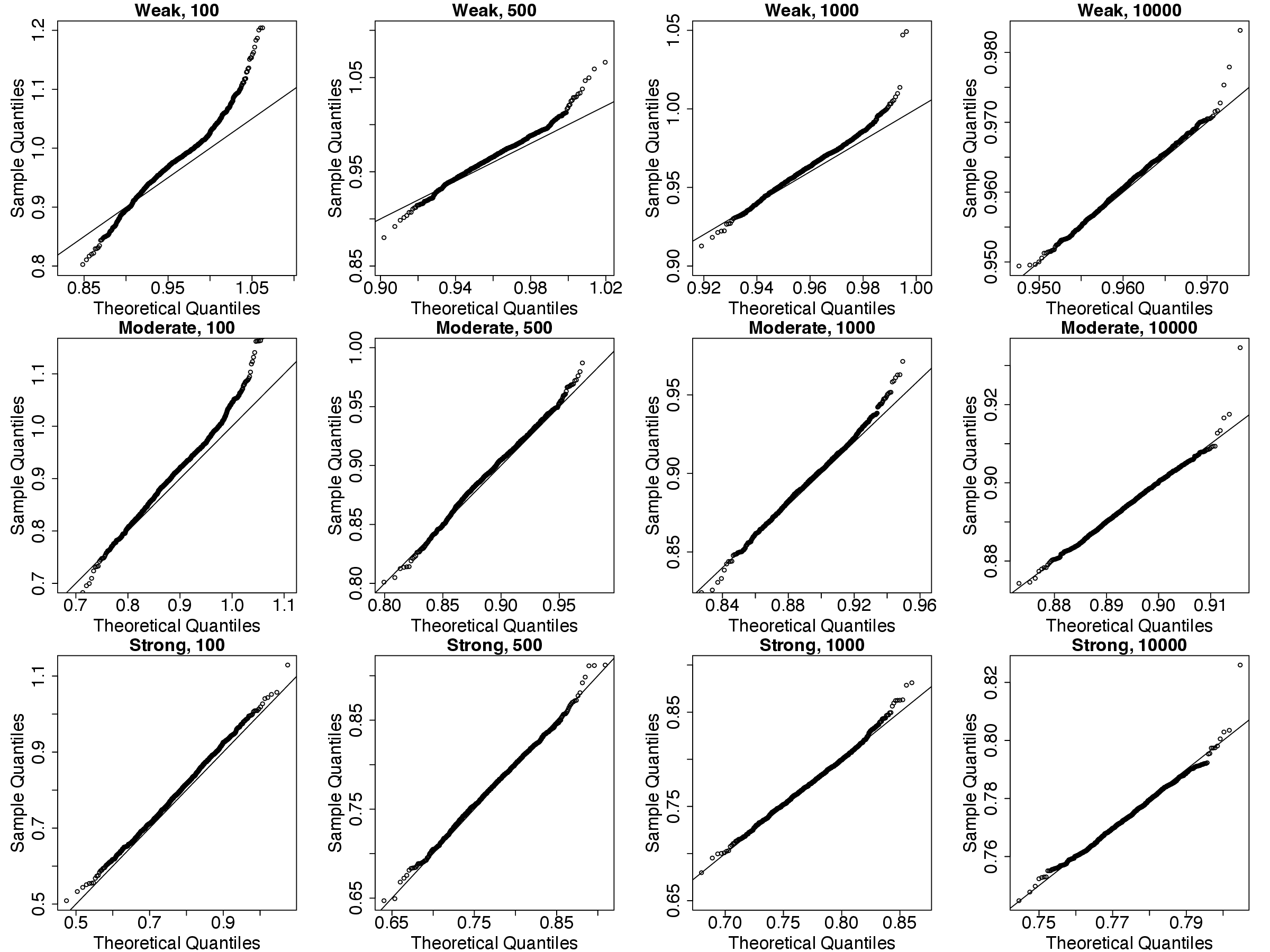}
	\caption{QQ plots of $\hatB_{01}$ against the distribution derived in Theorem~\ref{theo:nonind}.  The top row considers the case where $\sigma = 1.5$, the middle row has $\sigma = 2$, and the bottom row has $\sigma = 3$.  The columns correspond to $n \in \{100, 500, 1000, 10000\}$. 
    \label{fig:qq}}
\end{figure}

Figure~\ref{fig:qq} presents quantile-quantile (QQ) plots of the empirical distribution of $\hatB_{01}$ with $\sigma \in \{1.5, 2, 5\}$ based on simulated data of sizes $n \in \{100, 500, 1000, 10000\}$ against the theoretical distribution from Theorem~\ref{theo:nonind}.  Under moderate ($\sigma=2$) and strong ($\sigma=3$) dependence,  the distributions agree when $n \ge 500$, and the degree of agreement increases as the sample size increases in all scenarios, validating the asymptotic theory. Under weak ($\sigma = 1.5$) dependence, the empirical distribution has tails longer than a normal distribution when $n=100$. 
This is expected, as the estimator has a different asymptotic distribution under independence; recall Theorem~\ref{theo:ind}.
The  distribution of the estimator under independence is related to the integrated square error of KDEs. As discussed in \cite{li1996nonparametric}, with finite samples, the integrated square error approaches its asymptotic distribution slowly. 

Table~\ref{tab:simp} reports the empirical bias and standard deviation in the Bhattacharyya coefficient estimator \eqref{eq:split} and empirical coverage of a nominal 95\% confidence interval \eqref{eq:cib} based on 1000 independent replications when $\sigma \in \{1.5, 2, 3\}$ and $n\in \{500, 1000, 10000\}$. With consistently small bias and standard deviation, the proposed estimator, $\hatB_{01}$, exhibits good overall performance. As expected, the performance improves as the sample size increases. The standard deviation increases as dependence becomes stronger, consistent with the theory presented earlier. The coverage of the large-sample confidence interval is close to the nominal $95\%$ under moderate and high dependence, which is consistent with  Figure \ref{fig:qq}. Under weak dependence, a larger sample size is required for the correct coverage. 

\begin{table} \centering 
	\caption{Estimated bias and standard deviation of $\hatB_{01}$, and empirical coverage rate of nominal 95\% confidence intervals for $\B_{01}$ (all multiplied by 100) under three levels of dependence.  \label{tab:simp}}
	\begin{tabular}{@{\extracolsep{5pt}} cccc|ccc|ccc} 
		\toprule
		& \multicolumn{3}{c}{$n=500$}& \multicolumn{3}{c}{$n=1,000$}&\multicolumn{3}{c}{$n=10,000$}\\ 
		$\sigma$  & Bias & SD & Cov & Bias& SD& Cov & Bias & SD & Cov  \\ 
		\midrule
		1.5  & 0.56 & 2.18 & 80.3 & 0.39 & 1.77 & 85.2 & 0.05 & 0.42 & 93.5 \\ 
		2  & 0.38 & 3.02 & 91.1 & 0.27 & 2.53 & 91.6 & 0.01 & 0.66 & 95.4 \\ 
		3  & 0.22 & 3.91 & 95.5 & 0.17 & 3.17 & 94.1 & -0.05 & 0.90 & 97 \\
		\bottomrule
	\end{tabular} 
\end{table}

\begin{figure}
	\centering
	\includegraphics[width=.7\textwidth]{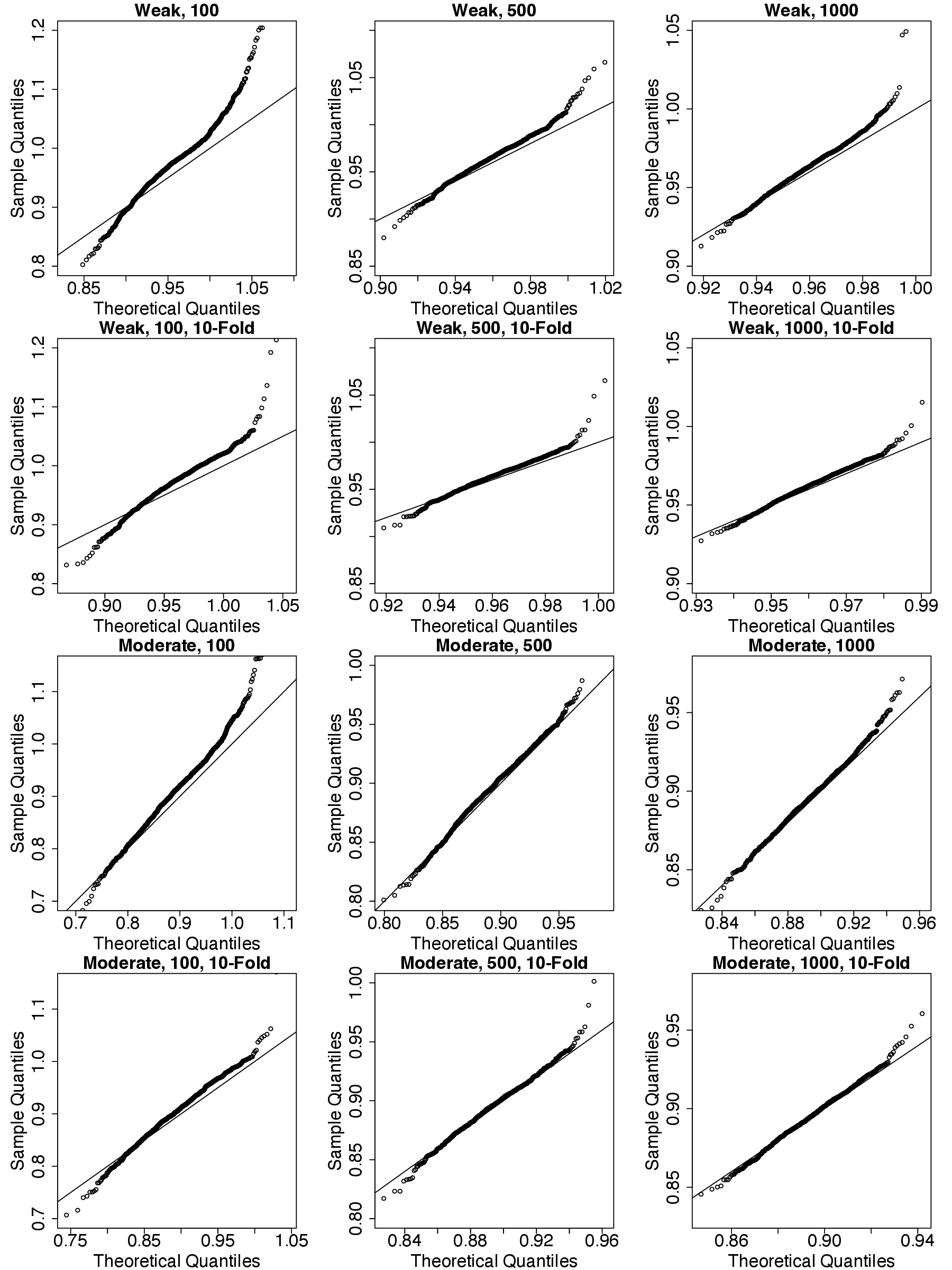}
	\caption{QQ plots of $\hatB_{01}$ (rows 1 and 3) and $\hatB_{01}^{10-fold}$ (rows 2 and 4).  Rows 1 and 2 correspond to weak ($\sigma=1.5$) dependence while rows 3 and 4 correspond to moderate ($\sigma=2$) dependence.  The columns correspond to $n \in 
    \{100, 500, 1000\}$. \label{fig:qqcv}}
\end{figure}

Recall from Proposition~\ref{prop:cv} that using $M$-fold estimation results in a smaller asymptotic variance.  Figure~\ref{fig:qqcv} presents QQ plots of the empirical distributions of $\hatB_{01}$ and the 10-fold cross-validated estimator, $\hatB_{01}^{10-fold}$, against the appropriate asymptotic distributions under weak ($\sigma=1.5$) and moderate ($\sigma=2$) dependence and when $n \in \{100,500,1000\}$. We omit the results for strong dependence, as they were already quite satisfactory. It is clear that the empirical distribution of the 10-fold cross-validated estimator aligns more closely with its asymptotic distribution across all scenarios.

Table~\ref{tab:simpcv} presents the results of two-fold, five-fold, and ten-fold estimation of the Bhattacharyya coefficient and the coverage of nominal 95\% confidence intervals. Comparing the results with Table ~\ref{tab:simp} demonstrates that these estimators result in smaller bias and standard deviation and improved coverage rates. Note that the confidence intervals are also narrower by a factor of $1/\sqrt{2}$.

\begin{table} \centering 
	\caption{Estimated bias and standard deviation of $\hatB_{01}^{2-fold}$, $\hatB_{01}^{5-fold}$, and $\hatB_{01}^{10-fold}$ and empirical coverage rate of nominal 95\% confidence intervals for $\B_{01}$ (all multiplied by 100) under three levels of dependence. \label{tab:simpcv}}
	\begin{tabular}{@{\extracolsep{5pt}} ccccc|ccc|ccc} 
		\toprule 
		&& \multicolumn{3}{c}{$n=500$}& \multicolumn{3}{c}{$n=1000$}&\multicolumn{3}{c}{$n=10000$}\\ 
		& $\sigma$ & Bias & SD & Cov & Bias& SD& Cov & Bias & SD & Cov  \\ 
\midrule
2-fold &	1.5 & 0.6 & 1.80 & $76.1$ & 0.33 & 1.25 & $83.7$ & 0.05 & 0.30 & $92.2$ \\ 
		&	2 & 0.37 & 2.35 & $88.9$ & 0.19 & 1.77 & $90.6$ & 0.01 & 0.48 & $94.4$ \\ 
		&	3 & 0.16 & 2.90 & $94.1$ & 0.06 & 2.23 & $94.1$ & -0.05 & 0.65 & $95.5$ \\ 
\midrule
5-fold&	1.5 & 0.38 & 1.53 & $83.3$ & 0.21 & 1.04 & $86.6$ & 0.03 & 0.30 & $93.7$ \\ 
		&	2 & 0.20 & 2.19 & $91.9$ & 0.10 & 1.57 & $92$ & -0.01 & 0.47 & $94.9$ \\ 
		&	3 & 0.03 & 2.81 & $95.4$ & -0.04 & 2.06 & $94.7$ & -0.05 & 0.65 & $96$ \\ 
		\midrule
		10-fold&	1.5 & 0.34 & 1.52 & $84.8$ & 0.20 & 1.03 & $87.8$ & 0.03 & 0.31 & $92.8$ \\ 
		&	2  & 0.18 & 2.20 & $92$ & 0.09 & 1.57 & $91.6$ & -0.00 & 0.49 & $94.6$ \\ 
		&	3 & 0.02 & 2.83 & 95 & -0.03 & 2.08 & 94 & -0.04 & 0.67 & 95.4 \\ 
\bottomrule	\end{tabular} 
\end{table}

\subsubsection{Independence test}
\label{sec:simuind}

\begin{figure} 
\centering
\includegraphics[width=.5\textwidth]{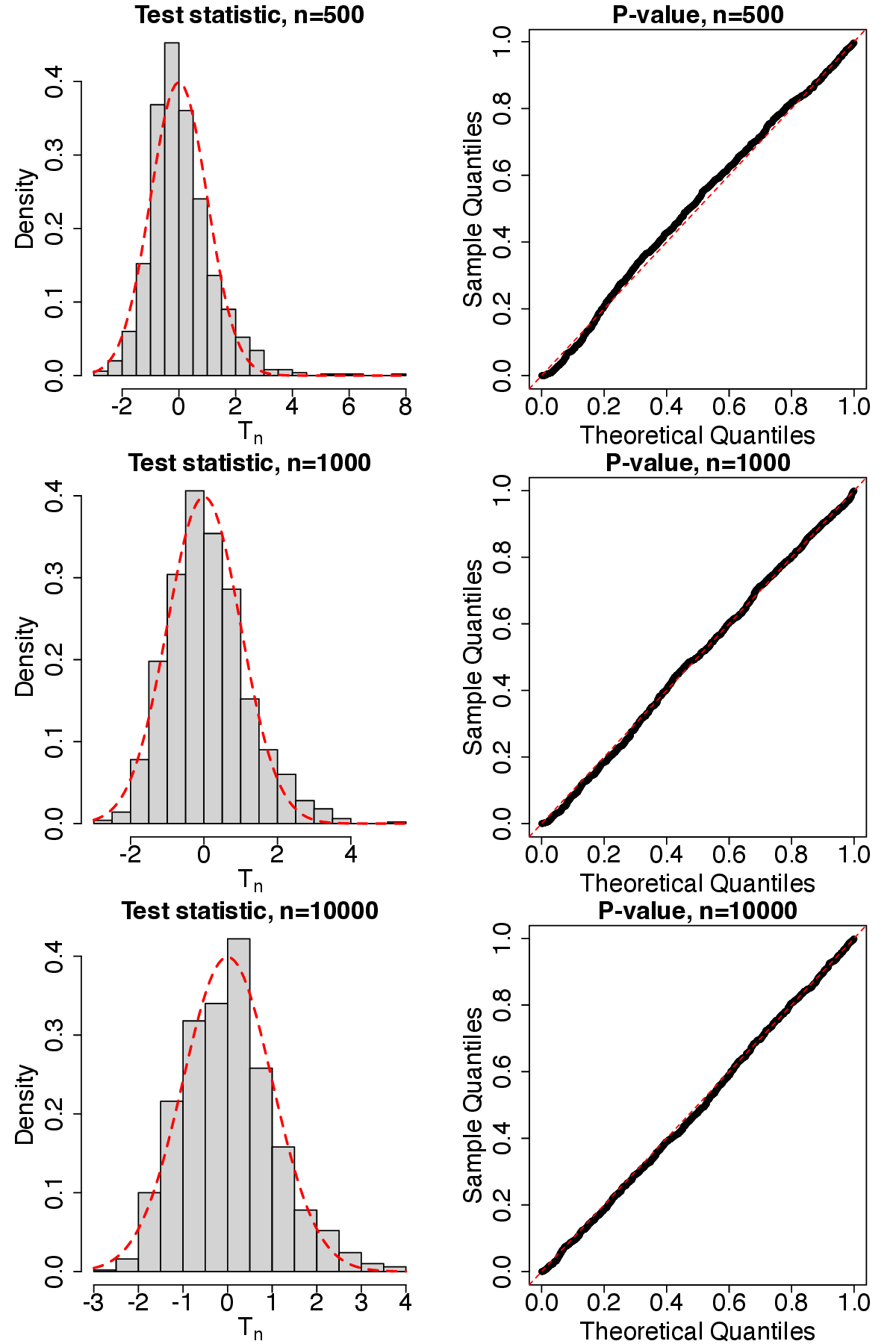}
\caption{Left column: Histograms of the test statistic compared with the theoretical distribution (dashed curve). Right column: QQ plots of p-values under independence. The sample sizes are 500, 1000, and 10000 in the top, middle, and bottom rows, respectively.\label{fig:qqtest}}
\end{figure}

Consider the proposed independence test in Section~\ref{sec:test}. Suppose $X \sim \mathrm{Bernoulli}(0.4)$ and $Y\sim \mathrm{Uniform}(0,1)$ independently. 
Based on 1000 replications for each of $n \in \{500, 1000, 10000\}$, Figure~\ref{fig:qqtest} presents a histogram of the test statistic and a QQ plot of the observed p-values under the null hypothesis.   In all three cases, the test statistic appears to follow a standard normal distribution and the p-value a uniform distribution, as expected from Theorem~\ref{theo:ind}. Recall that our test has a closed-form p-value, and thus no bootstrapping is required.

Table \ref{tab:power} presents estimates of the level and power of the test in the three settings from Section~\ref{sec:simuest} for $n \in \{500,1000, 10000\}$ based on 1000 replications. The significance level of the test was set to 0.05. The power of the proposed test increases with both the sample size and the strength of dependence. Overall, the test exhibits Type I error control under the null hypothesis and demonstrates good power against dependence. 

\begin{table}  \centering  \caption{Estimated level and power of the proposed independence test for the three dependence settings from Section~\ref{sec:simuest}.  \label{tab:power}} 
	\begin{tabular}{@{\extracolsep{5pt}} ccccc} 
			\toprule
		$n$	&Level& \multicolumn{3}{c}{Power}\\
		\cline{3-5} 

	&  & Weak & Moderate & Strong \\ 
	\midrule
500 & $0.074$ & $0.703$ & $0.990$ & $1$ \\ 
1000 & $0.071$ & $0.923$ & $1$ & $1$ \\ 
10000 & $0.048$ & $1$ & $1$ & $1$ \\ 
%		1000 & $0.078$ & $0.386$ & $1$ & $1$ \\ 
%		10000 & $0.048$ & $1$ & $1$ & $1$ \\ 
%		20000 & $0.052$ & $1$ & $1$ & $1$ \\ 
\bottomrule
	\end{tabular} 
\end{table} 

We address that independence testing is not our main focus, as mentioned in Section~\ref{sec:intro}. Since our independence test has closed-form p-values, it can be used as a tool for exploration at little computational cost.
There are many independence tests available for in-depth examination (e.g., \citealt{gretton2005kernel,berrett2019nonparametric}).

\section{Multinomial and Continuous Variables}
\label{sec:ext}
When the discrete variable has possible values $1,\ldots,J$ with $J > 2$, the dependence between $X$ and $Y$ can be quantified using the mutual similarity between $Y \mid X=1,~Y \mid X=2,\cdots,Y \mid X=J$. We utilize pairwise Hellinger distances and define
\begin{align*}
	\cH^2(f_{Y \mid X=1},\ldots,f_{Y \mid X=J})= & \frac{2}{J(J-1)}\left[\sum_{1\leq l < j \leq J}\mathcal{H}^2(f_{Y\mid X=l},f_{Y \mid X=j})\right]\\
    = & 1-\frac{2}{J(J-1)}\sum_{1\leq l < j \leq J}\int\sqrt{f_{Y \mid X=l}(y)f_{Y \mid X=j}(y)}\mathrm{d}y\\
    = & 1-\frac{2}{J(J-1)}\sum_{1 \leq l < j \leq J}\B \left(f_{Y \mid X=l},f_{Y \mid X=j}\right).
\end{align*}
Let $\B_{1J} =\sum_{1 \leq l < j \leq J}\B\left(f_{Y \mid X=l},f_{Y \mid X=j}\right)$ so that the proposed measure of dependence is
\begin{equation}
\label{eq:multinomial VH}
    \cV_{H}(X, Y) = \cH (f_{Y \mid X=1},\ldots,f_{Y \mid X=J})= \left[ 1-\frac{2}{J(J-1)}\B_{1J} \right]^{1/2} .
\end{equation}
Notice that the expression in \eqref{eq:multinomial VH}
 coincides with the expression in \eqref{eq:binary VH} when $J=2$. Moreover, it preserves  the desiderata that  that $0 \le \cV_H (X,Y) \le 1$ and $\cV_H(X,Y) = 0$ if and only if $X$ and $Y$ are independent, or equivalently the conditional distributions $f_{Y \mid X=1},\ldots,f_{Y \mid X=J}$ are all identical. In addition, $\cV_H (X, Y) = 1$ if and only if the supports of the conditional distributions $f_{Y \mid X=1},\ldots,f_{Y \mid X=J}$ do not overlap such that the value of $Y$ can perfectly determine $X$.  Finally, the nominal encoding of $X$ does not change $\cV_H(X,Y)$. 
 
Similarly to the development for binary outcomes, we propose a data-splitting estimator for $\B_{1J}$
\begin{align}\label{eq:split3}
	\begin{split}
        \hatB_{1J} = & \sum_{1 \leq l < j \leq J}\left[\frac{1}{2\sum_{i=\lfloor n/2 \rfloor +1}^n 1\left(X_i=l\right)} \sum_{i=\lfloor n/2 \rfloor +1}^n 1\left(X_i=l\right) \sqrt{\frac{\hat{f}_{Y \mid X=j}(Y_i)}{\hat{f}_{Y \mid X=l}(Y_i)}}\right.\\
        &\left.~~~~~~~\qquad+\frac{1}{2\sum_{i=\lfloor n/2 \rfloor +1}^n 1\left(X_i=j\right)}\sum_{i=\lfloor n/2 \rfloor +1}^n 1\left(X_i=j\right)\sqrt{\frac{\hat{f}_{Y \mid X=l}(Y_i)}{\hat{f}_{Y \mid X=j}(Y_i)}}\right].\end{split}
\end{align}

\subsection{Asymptotic Properties}\label{sec:theomulti}

Theorems~\ref{theo:nonindmulti} and~\ref{theo:indmulti} establish the asymptotic distribution of $\hatB_{1J}$ for multinomial data in the non-independent and independent cases, respectively, and hence generalize Theorems~\ref{theo:nonind} and~\ref{theo:ind}.
Assumption \ref{assum:1} is required for $f_{Y|X=j},1\leq j\leq J$.
Let $p_j=\Pr(X=j)$  for $j=1,\ldots,J$ and $\hat{p}_j$ be the usual empirical estimator.

\begin{theorem}
\label{theo:nonindmulti}
  Under Assumption \ref{assum:1},   if $X$ and $Y$ are not independent,  then, as $n \to \infty$,
    \begin{equation*}
        \sqrt{n} \left(\hatB_{1J} - \B_{1J} \right) \rightarrow_d N\left(0,V \right),
    \end{equation*}where 
        \begin{equation*}
        V = \sum_{j=1}^{J}\frac{J-1+\sum_{k,l\neq j,k<l}2\B\left(f_{Y \mid X=k},f_{Y \mid X=l}\right)-\left(\sum_{l\neq j}\B\left(f_{Y\mid X=l},f_{Y\mid X=j}\right)\right)^2}{2p_j}.
    \end{equation*}
\end{theorem}

Recall the definitions of $S_0$, $\sigma_{K}^{2}$, and $\kappa_{K}$ from Theorem~\ref{theo:ind}.
\begin{theorem}\label{theo:indmulti}
		If $X$ and $Y$ are  independent, then, as $n \to \infty$,
	\begin{align*}
	T_n=	\frac{	{\sqrt{n^2h}} \left[\hatB_{1J}-\frac{J(J-1)}{2}-\frac{\sigma_K^2 S_0 (J-1)}{4nh} \left(\sum_{j=1}^J\frac{1}{\hat{p}_j} \right) \right]}{\sqrt{S_0 \left\lbrace \left(\frac{\kappa_K^2 }{8}+\frac{\sigma_K^2}{4}\right) \left[\sum_{1 \leq l < j \leq J} \left(\frac{2}{\hat{p}_j \hat{p}_l}\right) + \sum_{j=1}^J \frac{(J-1)^2}{\hat{p}_j^2}\right]
				\right\rbrace}}\rightarrow_d N\left(0,1\right).
	\end{align*}
\end{theorem}
Similar to the binary setting, our theoretical results enable inference without resorting to simulation.
Letting
\begin{equation*}
        \hat{V} = \sum_{j=1}^{J}\frac{J-1+\sum_{k,l\neq j,k<l}2\hatB\left(\hat{f}_{Y \mid X=k},\hat{f}_{Y \mid X=l}\right)-\left(\sum_{l\neq j}\hatB\left( \hat{f}_{Y\mid X=l},\hat{f}_{Y\mid X=j}\right)\right)^2}{2\hat{p}_j},
    \end{equation*}
a $100(1-\alpha)\%$ confidence interval for $\B_{1J}$ is 
\begin{equation}
	\label{eq:multici}
    \hatB_{1J} \pm z_{1-\alpha/2} \left[ \frac{\hat{V}}{n} \right]^{1/2} .
\end{equation}
The confidence interval for $\cV_{H}(X, Y)$ can be obtained subsequently as a monotone function of $\B_{1J}$.
The p-value for the independence test can also be obtained straightforwardly as $2\left(1-\Phi\left(|T_n|\right)\right)$.

\subsection{Illustrative Examples}
\label{sec:simumulti}

\begin{figure} \centering
	\includegraphics[width=.9\textwidth]{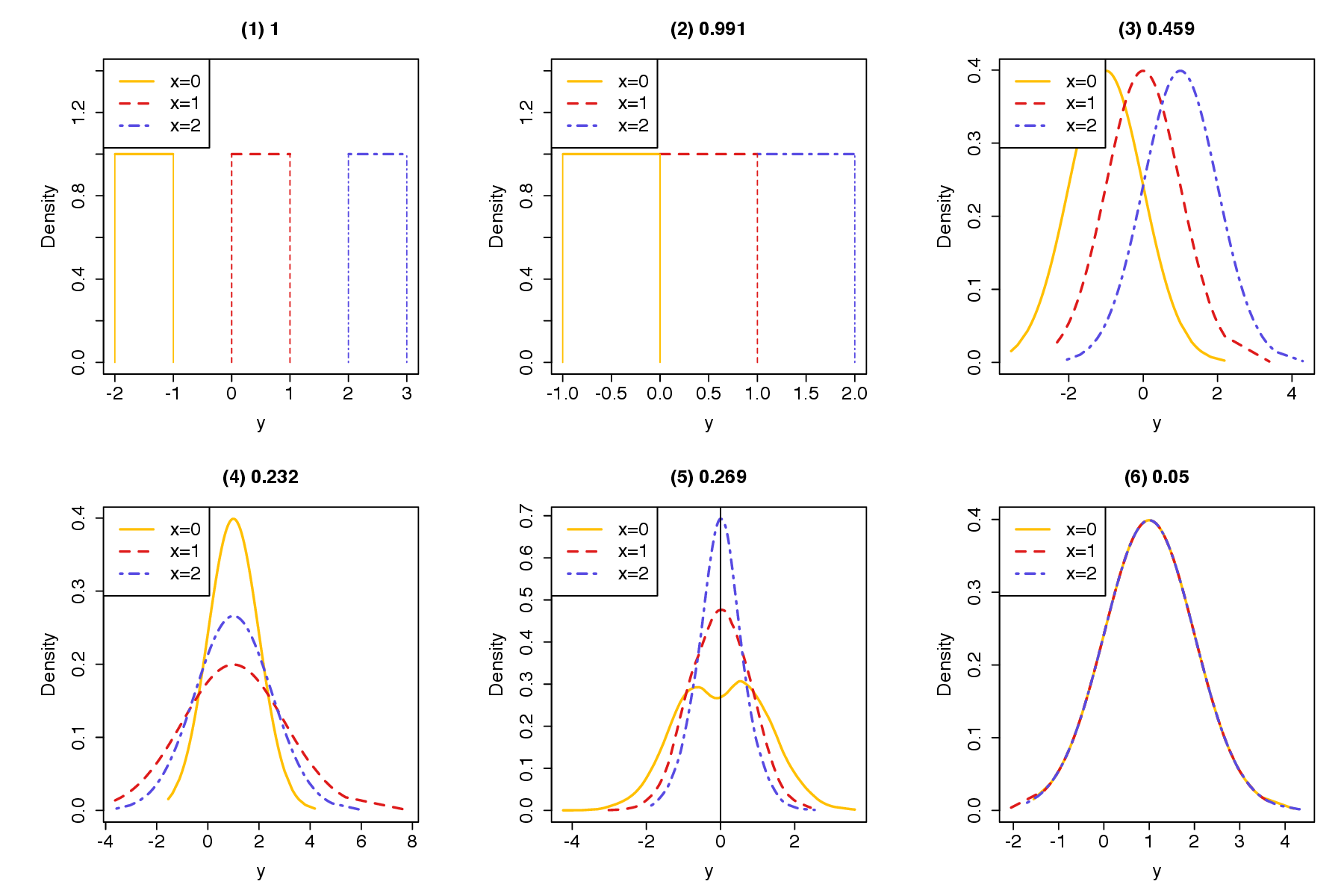}
	\caption{Values of $\hat{\mathcal{V}}_H(X, Y)$ %
    in six scenarios. Sample size 1000.\label{fig:demomulti}}
\end{figure}

Similar to Section~\ref{sec:simudemo}, the dependence measure $\cV_{H}(X,Y)$ will be considered in several scenarios, which we visualize in Figure~\ref{fig:demomulti}.
In scenarios 1 - 4 and 6, let $X \sim \mathrm{Multinomial}$ with levels $(0,1,2)$ and probabilities $(0.3, 0.3, 0.4)$.  Scenario 5 will be described below. 
\begin{enumerate}
	\item Suppose $Y \mid X=0 \sim \mathrm{Uniform}(-2,-1)$, $Y \mid X=1 \sim \mathrm{Uniform}(0,1)$, and $Y \mid X=2 \sim \mathrm{Uniform}(2,3)$.  This is a clear case of strict dependence in which the supports do not overlap and are well separated. 
	\item  Suppose $Y \mid X=0 \sim \mathrm{Uniform}(-1,0)$,  $Y \mid X=1 \sim \mathrm{Uniform}(0,1)$, and $Y \mid X=2 \sim \mathrm{Uniform}(1,2)$. Again, the supports do not overlap but are adjacent, resulting in a case of strict dependence; it is perhaps less clearly so than in the first scenario. 
	\item  Suppose $Y \mid X=0 \sim \mathrm{N}(0,1)$, $Y \mid X=1 \sim \mathrm{N}(-1,1)$, and $Y \mid X=2 \sim \mathrm{N}(1,1)$, which is intended to illustrate a difference in mean.
	\item Difference in variance. Suppose $Y \mid X=0 \sim \mathrm{N}(1,1)$,  $Y \mid X=1 \sim \mathrm{N}(1,2^2)$, and $Y \mid X=2 \sim \mathrm{N}(1,1.5^2)$, which is a case of a difference in variance.
	\item  Suppose $Y \sim \mathrm{N}(0,1)$ and $X \mid Y $ is distributed as a multinomial random variable with probabilities characterized by a piecewise multinomial logit regression.  Specifically, let 
    $D = 1+\exp\left(-2 Y1(Y<0)+2 Y1(Y>0)\right)+\exp\left(- Y1(Y<0)+ Y1(Y>0)\right),$
    \begin{align*}
        P(X=1 \mid Y) & =\frac{\exp\left(-2 Y1(Y<0)+2 Y1(Y>0)\right)}{D}\\
        P(X= 2 \mid Y)& =\frac{\exp\left(- Y1(Y<0)+ Y1(Y>0)\right)}{D}, 
    \end{align*}
    and $P(X=0 \mid Y) = 1 - P(X=1 \mid Y) - P(X=2 \mid Y)$.
    This is a case of nonlinear dependence. 
	\item Suppose $X$ is independent of $Y \sim \mathrm{N}(0,1)$.
    \end{enumerate}

Table~\ref{tab:othermulti} includes estimates of $\cV_{H}(X,Y)$ along with estimates of the  Pearson, Kendall,  Spearman, distance, and semi-distance correlation coefficients. 
Consistent with our observations in Section \ref{sec:simudemo},  $\hat{\mathcal{V}}_H(X, Y)$ is close to  1 under strict dependence in scenarios 1 and 2, contrary to other methods. In addition, it can detect various types of dependence in scenarios 3 - 5.

\begin{table} \centering 
	\caption{Estimated dependence measures in six scenarios with multinomial and continuous outcomes.
		\label{tab:othermulti} }
	\begin{tabular}{@{\extracolsep{5pt}} ccccccc} 
		\toprule
		Scenario& $\hat{\mathcal{V}}_H(X, Y)$ & Pearson & Kendall & Spearman & Distance &Semi-distance\\ 
\midrule 
1 & $1$ & $0.985$ & $0.809$ & $0.936$ & $0.988$ & $0.825$ \\ 
2 & $0.991$ & $0.944$ & $0.809$ & $0.936$ & $0.947$ & $0.730$ \\ 
3 & $0.459$ & $0.643$ & $0.521$ & $0.650$ & $0.625$ & $0.296$ \\ 
4 & $0.232$ & $0.003$ & $0.011$ & $0.014$ & $0.113$ & $0.019$ \\ 
5 & $0.269$ & $$-$0.010$ & $$-$0.012$ & $$-$0.015$ & $0.198$ & $0.029$ \\ 
6 & $0.050$ & $0.008$ & $0.006$ & $0.008$ & $0.031$ & $0.001$ \\ 
\bottomrule
	\end{tabular} 
\end{table} 

Furthermore, when the discrete variable is nominal, the encoding of its levels should not affect the value of the dependence measure. Above, the results were based on encoding the levels of $X$ as $(0,1,2)$.  Table~\ref{tab:othermulticode} presents the results of, instead, encoding the levels of $X$ as $(0,2,1)$. Pearson, Kendall, Spearman, and distance correlations are sensitive to the encoding of the discrete variable, while the values of $\hatcV_{H}(X,Y)$ and the semi-distance correlation do not change from Table~\ref{tab:othermulti}.  Under perfect dependence in scenario 1, all the other dependence measures are significantly smaller than 1 after the levels are shuffled.

\begin{table} \centering 
	\caption{Estimated dependence measures in six scenarios with multinomial and continuous outcomes when the multinomial is based on an alternative encoding.
		\label{tab:othermulticode} }
	\begin{tabular}{@{\extracolsep{5pt}} ccccccc} 
		\toprule
		Scenario& $\hat{\mathcal{V}}_H(X, Y)$ & Pearson & Kendall & Spearman & Distance&Semi-distance \\ 
		\midrule
        1 & $1$ & $0.473$ & $0.239$ & $0.421$ & $0.812$ & $0.825$ \\ 
2 & $0.991$ & $0.453$ & $0.239$ & $0.421$ & $0.757$ & $0.730$ \\ 
3 & $0.459$ & $0.290$ & $0.206$ & $0.292$ & $0.468$ & $0.296$ \\ 
4 & $0.232$ & $$-$0.020$ & $$-$0.0003$ & $$-$0.001$ & $0.155$ & $0.019$ \\ 
5 & $0.269$ & $$-$0.006$ & $$-$0.009$ & $$-$0.012$ & $0.168$ & $0.029$ \\ 
6 & $0.050$ & $$-$0.025$ & $$-$0.019$ & $$-$0.025$ & $0.036$ & $0.001$ \\ 
\bottomrule	\end{tabular} 
\end{table} 

\subsection{Simulation Experiment}
\label{sec:mutli_sim}

Now consider extending the setting from Section~\ref{sec:simu} to the multinomial setting with a moderate number of levels in the discrete outcome.  Specifically, assume $X$ follows a multinomial distribution with 5 levels, each with a probability of 0.2 and, for $j=1,\ldots,5$, $Y \mid X=j \sim \mathrm{N}(1,1+(j-1)\sigma)$, where $\sigma \in \{0.5, 1, 2\}$.  Note that $\sigma=0.5$ corresponds to weak dependence, $\sigma=1$ to moderate dependence, and $\sigma=2$ to strong dependence.

Table~\ref{tab:simp5} reports the empirical bias and standard deviation in estimating the Bhattacharyya coefficient \eqref{eq:split3} and empirical coverage of a nominal 95\% confidence interval \eqref{eq:multici}, based on 1000 independent replications for $n\in \{500, 1000, 10000\}$. It shows the good performance of the proposed estimator with moderate levels in the discrete outcome 
despite smaller sample sizes allocated for each category. 
When $J$ becomes large, it will require that $n/J$ remain approximately constant to ensure comparable sample sizes across groups, as argued in \cite{zhan2014testing}.

\begin{table} \centering 
	\caption{Bias, empirical standard deviation of $\hatB_{1J}$ and coverage rate of a 95\% large sample confidence interval. All results for Bias, SD, and Cov are multiplied by 100.  \label{tab:simp5}}
	\begin{tabular}{@{\extracolsep{5pt}} cccc|ccc|ccc} 
		\toprule
		& \multicolumn{3}{c}{$n=500$}& \multicolumn{3}{c}{$n=1,000$}&\multicolumn{3}{c}{$n=10,000$}\\ 
		 $\sigma$ & Bias & SD & Cov & Bias& SD& Cov & Bias & SD & Cov  \\ 
		\midrule
		$0.5$ & $8.927$ & $21.336$ & $85.4$ & $5.249$ & $19.697$ & $88.8$ & $1.044$ & $5.130$ & $93.5$ \\ 
		1 & $7.070$ & $23.352$ & $93.4$ & $4.129$ & $21.571$ & $93.9$ & $0.663$ & $5.944$ & $95.5$ \\ 
		 2& $5.926$ & $24.170$ & $96.1$ & $3.407$ & $22.213$ & $96$ & $0.427$ & $6.162$ & $97.5$ \\ 		
		\bottomrule
	\end{tabular} 
\end{table}

\section{Applications}
\label{sec:applications}
The use of the proposed dependence measure is illustrated with two data examples.  In Section~\ref{ex:redspiral}, the dependence measure quantifies the association between the amount of ongoing star formation in a galaxy and that galaxy's bulge size. In Section~\ref{sec:insurance}, it provides a measure of association between auto insurance claim amount and car models. 

\subsection{Spiral Galaxies}
\label{ex:redspiral}

\begin{figure}
	\centering
        \includegraphics[width=.6\textwidth]{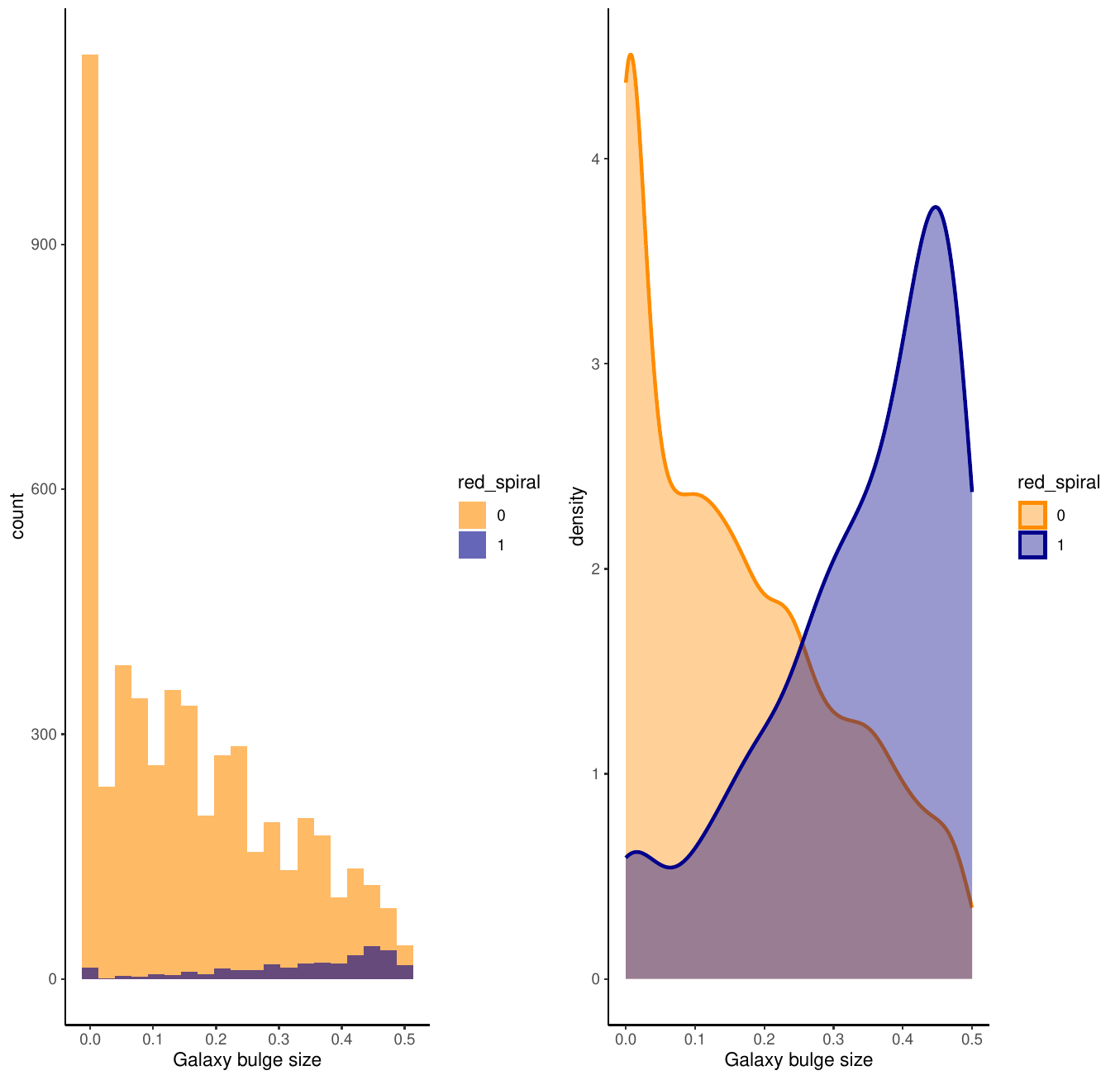}
	\caption{Histograms of counts and estimated densities of red spiral ($\texttt{red\_spiral}=1$) and blue spiral ($\texttt{red\_spiral}=0$) galaxies.\label{fig:red_spiral}}
\end{figure}

The rest-frame color of a spiral galaxy is associated with the amount of
star formation, with red having more and blue having less.  \citet{mast:2010} present data for a total of 5433 galaxies in two categories: 294 red
spirals and 5139 blue spirals. 
The galaxy bulge is a cluster of stars at its center, which is suspected to contain a supermassive black hole.  \citet{mast:2010} also provides a measure for bulge size based on the de Vaucouleurs profile, which is a continuous variable.  The data are displayed in Figure~\ref{fig:red_spiral}. The estimated dependence between galaxy bulge size and rest frame color is $\hatcV_H = 0.385$, and a 90\% confidence interval for $\cV_H$ is $(0.348, 0.422)$, which provides evidence of association between galaxy bulge size and rate of star formation.

\subsection{Insurance Claims}\label{sec:insurance}
In insurance practice, understanding the relationship between claim amounts and risk factors is a key step in risk analysis, as it can facilitate the practice of pricing and reserving. The Swedish Committee on the Analysis of Risk Premium in Motor Insurance compiled data describing $n=1797$ third-party automobile insurance claims for the year 1977 \citep{hallin1983swedish}. The interest here is in the relationship between the total payment and the type of vehicle. Vehicle model is a categorical variable with nine levels   coded from 1 to 9, while total payment is treated as a continuous variable.  
 
\begin{figure} \centering
	\includegraphics[width=.35\textwidth]{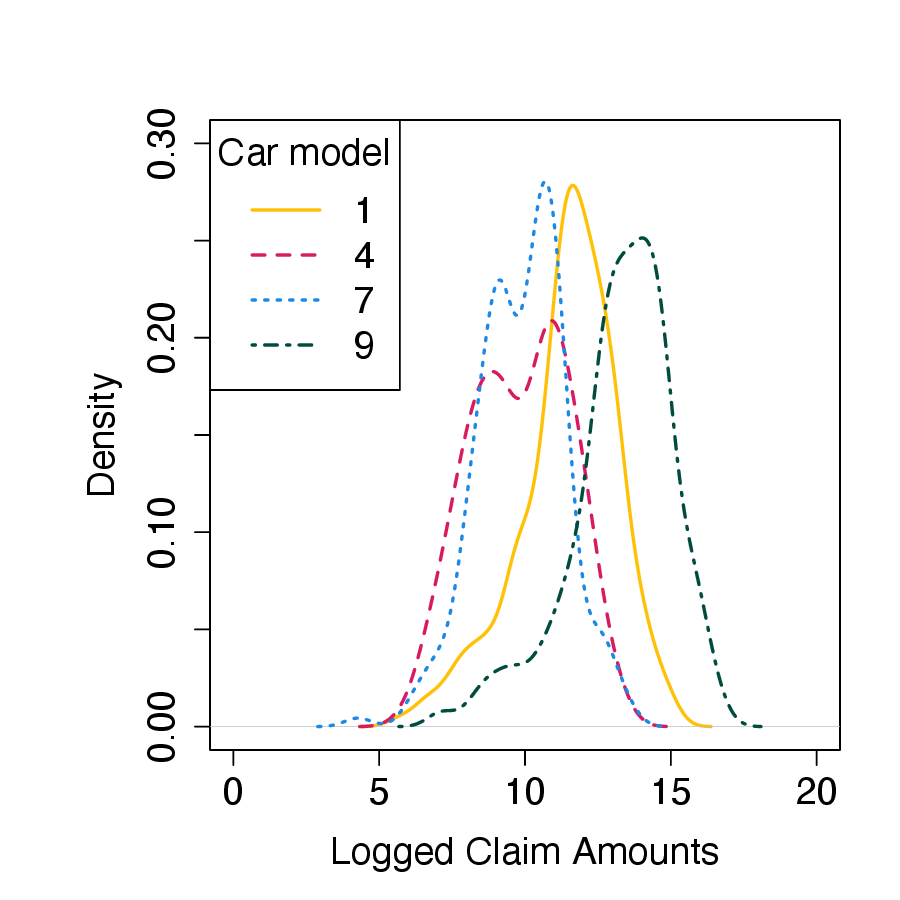}
	\caption{Density of total insurance claim by vehicle type. \label{fig:insurance}}
\end{figure}

 Table~\ref{tab:entity}  summarizes the distribution of the vehicle model, and Figure~\ref{fig:insurance} demonstrates the distribution of total insurance claim by vehicle type for a subset of the categories. It is apparent that vehicle type impacts the distribution of claim amounts.  Indeed, the estimated dependence between vehicle type and claim amount is $\hatcV_{H}(X,Y)=0.283$, and a 90\% confidence interval for $\cV_{H}(X,Y)$ is $(0.265, 0.3)$, indicating moderate and statistically significant dependence.
 
\begin{table} \centering 
	\caption{Distribution of vehicle model.
	\label{tab:entity} }
\begin{tabular}{@{\extracolsep{5pt}} ccccccccc} 
	\toprule
	1 & 2 & 3 & 4 & 5 & 6 & 7 & 8 & 9 \\ 
	\midrule
	$227$ & $205$ & $178$ & $155$ & $206$ & $212$ & $197$ & $173$ & $244$ \\ 
	\bottomrule
	\end{tabular} 
\end{table} 

\section{Final Remarks}
\label{sec:final}

We propose a new dependence measure for mixed continuous and multinomial outcomes. The dependence measure is based on the Hellinger distance for the continuous variable across subpopulations defined by the values of the multinomial variable and satisfies all the ideal properties. An estimator of the dependence measure based on data splitting is proposed. The theoretical properties of the estimator are studied, which has a simple asymptotic distribution and $\sqrt{n}$ convergence rate. One can hence construct a confidence interval and conduct an independence test straightforwardly without resorting to the bootstrap. 

The proposed dependence measure has many potential applications. For example, it can potentially serve as a tool for clustering validations. A clustering algorithm that results in a strong dependence between cluster labels and features is superior. In addition, it can be used for marginal feature screening, which filters out uninformative variables before applying standard penalized variable selection methods. Because our dependence measure is generally applicable across data types without distributional assumptions, it provides a versatile technique for feature screening in many scenarios with diverse response and feature types, in a model-free manner.

Extending the dependence measure to random vectors is an interesting topic for future work.  
We employ the KDE in the univariate case, which might become suboptimal for vectors. Methods of estimating  high-dimensional densities  \citep{nguyen2022adaptive}
or ratios of densities
\citep{sugiyama2010conditional} 
might be a remedy.

\appendix

\section{Appendix}\label{sec:proof}

%\section{Basic Properties of Population  \texorpdfstring{$\mathcal{V}_H(X,Y)$}{VH(X,Y)}}
%\label{app:basic props}
    \begin{lemma}
	$\mathcal{V}_H(X,Y)=0$ if and only if $X\ind Y.$ 
\begin{comment}
    
	\begin{proof}
		$\mathcal{H}(f_{Y|X=0},f_{Y|X=1})=0$ if and only of for $Y|X=0$  and $Y|X=1$ are identically distributed. This is  equivalent to the independence between $X$ and $Y$. 
		
	\end{proof}

\end{comment}
\end{lemma}

\begin{lemma}
	
	$\mathcal{V}_H(X,Y)=1$ iff and only if there is a strict dependence between $X$ and $Y$.
	\begin{comment}
	\begin{proof}
		The maximum distance 1 is achieved when the support of $f_{Y|X=0}$ assigns probability zero to every set to which $f_{Y|X=1}$ assigns a positive probability, and vice versa. %Equivalently, the value of $X$ determines the range of $Y$.
		
	\end{proof}
    \end{comment}
\end{lemma}

These two lemmas above are immediate consequences of the properties of the Hellinger distance. The proof of Lemma \ref{lemma:one} is included in the supplementary material.

\begin{lemma}\label{lemma:one}
	$\mathcal{V}_H(X,g(Y))=\mathcal{V}_H(X,Y)$ for a one-to-one measurable function $g$.
\end{lemma}

%\section*{Supplementary material}

%The supplementary material includes  proofs of the theoretical results in Sections \ref{sec:theo} and  \ref{sec:theomulti}. 

%\section*{Data Availability Statement}
%The data that support the findings of this study are available  at \url{https://instruction.bus.wisc.edu/jfrees/jfreesbooks/Regression%20Modeling/BookWebDec2010/data.html}.

\bibliographystyle{apalike}
 \bibliography{mixed}
\end{document}